\documentclass[hyper,letterpaper,12pt]{article}
\usepackage{a4wide}
\usepackage{amsmath}
\usepackage[
      colorlinks=true,
      linkcolor=blue,
      urlcolor=blue,
      filecolor=black,
      citecolor=blue,
            ]{hyperref}
\usepackage{color}
\usepackage{graphicx}
\usepackage{amsfonts}
\usepackage{amssymb}
\usepackage{epstopdf}

\def\beq{\begin{equation}}
\def\eeq{\end{equation}}

\begin{document}
\title{\bf \Large  Vector Condensate and AdS Soliton Instability Induced by a Magnetic Field }

\author{\large
~Rong-Gen Cai$^1$\footnote{E-mail: cairg@itp.ac.cn}~,
~~Li Li$^1$\footnote{E-mail: liliphy@itp.ac.cn}~,
~~Li-Fang Li$^2$\footnote{E-mail: lilf@itp.ac.cn}~,
~~You Wu$^3$\footnote{E-mail: 5u@ruc.edu.cn}\\
\\
\small $^1$State Key Laboratory of Theoretical Physics,\\
\small Institute of Theoretical Physics, Chinese Academy of Sciences,\\
\small Beijing 100190,  China.\\
\small $^2$State Key Laboratory of Space Weather, \\
\small Center for Space Science and Applied Research, Chinese Academy of Sciences,\\
\small Beijing 100190, China.\\
\small $^3$Department of physics, Renmin University of China, Beijing 100872, China.}
\date{\today}
\maketitle

\begin{abstract}
\normalsize We continue to study the holographic p-wave superconductor model in the
Einstein-Maxwell-complex vector field theory with a non-minimal coupling between the
complex vector field and the Maxwell field. In this paper we work in the AdS
soliton background which describes a conformal field theory in the
confined phase and focus on the probe approximation. We find that an applied
 magnetic field can lead to the condensate of the vector field and the AdS soliton
 instability. As a result, a vortex lattice structure forms in
the spatial directions perpendicular to the applied magnetic field.
As a comparison, we also discuss the vector condensate in the
Einstein-SU(2) Yang-Mills theory and find that in the setup of the
present paper, the Einstein-Maxwell-complex vector field model is a
generalization of the SU(2) model in the sense that the
vector field has a general mass and gyromagnetic ratio.

\end{abstract}

\tableofcontents

\section{ Introduction}
The gauge/gravity
duality~\cite{Maldacena:1997re,Gubser:1998bc,Witten:1998qj} allows
us to map a strongly interacting quantum field theory to a weakly
coupled gravity theory living in a higher dimensional spacetime. The
applications of the gauge/gravity duality have been extended to various
fields in recent years, especially in condensed matter
physics~\cite{Hartnoll:2009sz,Herzog:2009xv,McGreevy:2009xe}.

A well known example is the holographic superconductor
model~\cite{Hartnoll:2008vx,Hartnoll:2008kx}. By introducing a
charged scalar field coupled to a $(3+1)$ dimensional
Einstein-Maxwell theory with a negative cosmological constant, this
model admits a second order phase transition.  At high temperature,
the scalar field vanishes and one obtains the standard AdS
Reissner-Nordstr\"om black hole. However, when one lowers the
temperature, the AdS Reissner-Nordstr\"om black hole becomes
unstable and is replaced by a charged black hole with a non-trivial
scalar ``hair". The higher temperature AdS Reissner-Nordstr\"om
black hole describes a conductor phase, while the low temperature
black hole solution has the expected behavior for a superconducting
phase. Since the condensed field is a scalar field dual to a scalar
operator in the field theory side, it is therefore a holographic
s-wave superconductor model. Holographic p-wave models were realized by the
condensation of a vector like
field~\cite{Gubser:2008wv,Aprile:2010ge,Cai:2013pda,Cai:2013aca},
while holographic d-wave models were built up by introducing a
charged spin two field propagating in the
bulk~\cite{Chen:2010mk,Benini:2010pr}.

On the other hand, some novel properties of QCD matter in a strong
magnetic field have been revealed recently, see
ref.~\cite{Kharzeev:2012ph} for a review. In particular, the QCD
vacuum may undergo a phase transition to an exotic phase with
charged $\rho$-meson condensed in a sufficiently strong magnetic
field, which is a kind of anisotropic superconducting
phase~\cite{Chernodub:2010qx,Chernodub:2011mc,Chernodub:2011gs}.
Similar phenomenon has been also studied by holography based on the
Sakai-Sugimoto
model~\cite{Callebaut:2011ab,Callebaut:2013ria,Callebaut:2013wba}.
In particular, the holographic $\rho$-meson studied in the confined
phase at zero temperature~\cite{Callebaut:2011ab} suggests that the
effective mass of the $\rho$-meson would become tachyonic in a
sufficiently strong magnetic field, which gives rise to the
condensate of the $\rho$-meson in a strong magnetic field.  In
principle, the holographic p-wave model proposed in
refs.~\cite{Cai:2013pda,Cai:2013aca} can also describe the condensate of
the $\rho$-meson in low energy QCD.

In a recent paper~\cite{Cai:2013pda}, we constructed a holographic
model with vector condensate in the Einstein-Maxwell-complex vector
field theory in $(3+1)$ dimensional spacetime. In this model there
exists a non-minimal coupling between the complex vector field and
the Maxwell field. Working in the probe limit, we found that there
is a critical temperature below which the charged vector condenses
via a second order phase transition. The DC conductivity becomes
infinite and the AC conductivity develops a gap in the condensed
phase. In addition, it was found that the background magnetic field
can induce the condensate of the vector field even in the case
without chemical potential/charge density. In the case with
non-vanishing charge density, the transition temperature raises with
the applied magnetic field, and the condensate of the charged vector
operator forms a vortex lattice structure in the spatial directions
perpendicular to the applied magnetic field. Thus, this model is also
relevant to the $\rho$-meson condensate in an applied
magnetic field. In a follow up paper~\cite{Cai:2013aca} we studied
the model in some detail by taking into account the back reaction of
matter fields on a black hole background  geometry. Depending on
the mass and charge of the vector field, we found that not only
second order, but also first order and zeroth order phase
transitions exist in this model. In particular the so-called
``retrograde condensation" can also appear, there the hairy black
hole solution exists only for the temperature above a critical
value, its free energy is much larger than the one of the black hole without
the vector hair. For details please see ref.~\cite{Cai:2013aca}.

In the previous studies, we limited ourselves to the black hole
background, which corresponds to a dual field theory in the
deconfined phase in a finite temperature, according to the
gauge/gravity duality. It is worth generalizing our study to the AdS
soliton background case dual to a field theory in the confined phase. Such a study  can
not only shed lights on the p-wave insulator/superconductor
transition at zero temperature, but also is relevant to the
condensate of $\rho$-meson appearing in a confined phase. In this
model the background magnetic field is from the U(1) gauge field, so
we shall call it Abelian magnetic field for convenience. The black
hole instability induced by a background non-Abelian magnetic field
coming from a SU(2) gauge field has been reported in
refs.~\cite{Bu:2012mq,Wong:2013rda} where a vortex lattice structure
forms due to the condensation of some non-Abelian current operators.
To distinguish our model with the p-wave model with SU(2) gauge
field~\cite{Gubser:2008wv}, we may call our model Abelian p-wave
model, while the latter non-Abelian one.  To compare two models, we
will also investigate the AdS soliton instability induced by
non-Abelian magnetic field in the non-Abelian p-wave model.

 To immerse the strongly coupled system into an external magnetic
field background, we turn on a uniform magnetic field $B$ in the
bulk. For the Abelian p-wave model, we find that the increase of the
Abelian magnetic field will induce the instability of the AdS soliton
background. It is well known that the increase of the chemical
potential can induce the instability of the AdS soliton background,
which mimics the insulator/superconductor transition in many
holographic
s-wave~\cite{Nishioka:2009zj,Horowitz:2010jq,Peng:2011gh,Cai:2012es}
and non-Abelian p-wave~\cite{Akhavan:2010bf,Cai:2013oma} models. The
magnetic field here plays the same role as the chemical potential
and a family of condensate can be induced by the applied magnetic
field in the dual strongly coupled system. Although there is a tower
of ``droplet" solutions in the sense that they are localized in a
finite region for finite magnetic field~\cite{Albash:2008eh}, it is
quite different from the s-wave soliton case~\cite{Cai:2011tm},
where the presence of the magnetic field causes the phase
transition much more difficult. We then study the non-Abelian p-wave
model in the AdS soliton background with constant magnetic field.
Similarly, the non-Abelian magnetic field can also induce the AdS
soliton instability. In particular, if we take the mass of the
charged vector field to be zero and the non-minimal coupling
constant between the charged vector field and the Maxwell gauge
field to be one, we find that in the setup of the present paper, the reduced
equations of motion of the Abelian p-wave model are exactly the same
as those of the non-Abelian model. In this sense, the
Einstein-Maxwell-complex vector field model can be viewed as a
generalization of the SU(2) p-wave model. We also manage to
construct the vortex lattice solution in the condensed phase near
the transition point, which is very reminiscent of the Abrikosov
lattice in the common type-II superconductors.

The plan of this paper is as follows. In
section~\ref{sect:abeliabmodel}, we first introduce the holographic
model with a complex vector field charged under a U(1) gauge field
in the AdS soliton background. We turn on a uniform magnetic field
and give details about how to recover the Landau levels of the vector field
in this holographic model. Then we study the phase diagram. In
section~\ref{sect:nonabeliabmodel}, we investigate the magnetic
field effect for the non-Abelian p-wave model in the AdS soliton
background. We construct the general vortex lattice solutions in
section~\ref{sect:vortex} for both two models. The conclusion and
further discussions are included in section~\ref{sect:conclusion}.

\section{Einstein-Maxwell-complex vector model}
\label{sect:abeliabmodel}
In this section, we study the AdS soliton instabilities induced by an applied
magnetic field in the Einstein-Maxwell-complex vector field theory. Let us introduce a charged vector
field into a $(4+1)$ dimensional Einstein-Maxwell theory with a
negative cosmological constant. The full action reads
\begin{equation}\label{action}
S =\frac{1}{2\kappa^2}\int d^5 x
\sqrt{-g}[\mathcal{R}+\frac{12}{L^2}-\frac{1}{4}F_{\mu\nu} F^{\mu \nu}-\frac{1}{2}\rho_{\mu\nu}^\dagger\rho^{\mu\nu}-m^2\rho_\mu^\dagger\rho^\mu+iq\gamma \rho_\mu\rho_\nu^\dagger F^{\mu\nu}],
\end{equation}
where $\kappa^2\equiv 8\pi G$ is associated to the gravitational constant in
the bulk and $L$ is the AdS radius which is chosen to be unity. $m$ and $q$ are related to the mass and charge of the vector field $\rho_\mu$, respectively.
$F_{\mu\nu}=\nabla_\mu A_\nu-\nabla_\nu A_\mu$ is the strength of U(1) field $A_\mu$. The tensor $\rho_{\mu\nu}$
 is defined by $\rho_{\mu\nu}=D_\mu\rho_\nu-D_\nu\rho_\mu$ with the covariant
derivative $D_\mu=\nabla_\mu-iq A_\mu$. The last term in (\ref{action}) describes the
non-minimal coupling between the charged vector field $\rho_\mu$ and
U(1) gauge field $A_\mu$ characterizing the magnetic moment of the
vector field $\rho_\mu$. The coupling constant $\gamma$ in some
sense can be thought of as an effective gyromagnetic ratio of the
charged vector field.

The equation of motion for the charged vector field $\rho_\mu$ is given by
\begin{equation}\label{vector}
D^\nu\rho_{\nu\mu}-m^2\rho_\mu+iq\gamma\rho^\nu F_{\nu\mu}=0,
\end{equation}
while the equation of motion for gauge field $A_\mu$ is
\begin{equation}\label{gauge}
\nabla^\nu F_{\nu\mu}=iq(\rho^\nu\rho_{\nu\mu}^\dagger-{\rho^\nu}^\dagger\rho_{\nu\mu})+iq\gamma\nabla^\nu(\rho_\nu\rho_\mu^\dagger-\rho_\nu^\dagger\rho_\mu).
\end{equation}
By taking the limit $q\rightarrow\infty$ with $q\rho_\mu$ and $q
A_\mu$ fixed, the back reaction of the matter sources to the
background spacetime can be ignored. This is the probe limit we will adopt in
this paper. Since we want to study the phase transition in the
confined phase, the background geometry is taken to be a $(4+1)$-dimensional
AdS soliton
\begin{equation}\label{metric}
ds^2=\frac{dr^2}{f(r)}+r^2(-dt^2+dx^2+dy^2)+f(r)d\eta^2,
\end{equation}
where $f(r)=r^2(1-\frac{r_0^4}{r^4})$ and $r_0$ is the tip of the
soliton. This is a self-consistent solution of action (\ref{action})
by neglecting the matter sources. To avoid the potential conical
singularity at $r=r_0$, the spatial direction $\eta$ must have a
period $\eta\sim\eta+\pi/r_0$. The AdS soliton just looks like a
cigar with the asymptotical geometry $R^{1,2}\times S^1$ near the
AdS boundary. Since the spacetime exists only for $r>r_0$, the
dual field theory is in a confined phase and has a mass gap, $E_g \sim r_0$. The AdS soliton spacetime
can describe an insulator in condensed matter
physics~\cite{Nishioka:2009zj}, due to the presence of the mass gap.

\subsection{Turning on a constant magnetic field}
\label{sect:ansatzB}
We now turn on a magnetic field to study how the applied magnetic
field influences  the system. Similar to the  black hole case
studied in ref.~\cite{Cai:2013pda}, a consistent ansatz turns out to be
\begin{equation}\label{matterB}
\begin{split}
\rho_\nu dx^\nu=[\epsilon\rho_x(r,x)e^{ipy}+\mathcal{O}(\epsilon^3)]dx+[\epsilon\rho_y(r,x)e^{ipy}e^{i\theta}+\mathcal{O}(\epsilon^3)]dy,\\
A_\nu dx^\nu=[\phi(r)+\mathcal{O}(\epsilon^2)]dt+[Bx+\mathcal{O}(\epsilon^2)]dy,
\end{split}
\end{equation}
where $\rho_x(r,x)$, $\rho_y(r,x)$ and $\phi(r)$ are all real
functions, $p$ is a real constant and the constant $\theta$  is the
phase difference between the $x$ and $y$ components of the vector
field $\rho_{\mu}$. The constant Abelian magnetic field $B$ is
perpendicular to the $x-y$ plane. The parameter $\epsilon$ is a
small quantity characterizing the deviation from the critical point at
which the condensate begins to appear.

Substituting the ansatz~\eqref{matterB} into equation~\eqref{gauge},
the equation of motion for $\phi$ can be read off at the zeroth
order
\begin{equation}\label{eomphi}
\phi''(r)+(\frac{f'}{f}+\frac{1}{r})\phi'(r)=0.
\end{equation}
The asymptotic value of $\phi$ gives the chemical potential $\mu=A_t(r\rightarrow\infty)$ of the dual field theory. Imposing the Neumann-like boundary condition at the tip of the AdS soliton background~\cite{Nishioka:2009zj}, we find a unique solution
\begin{equation}\label{phi}
\phi(r)=\mu.
\end{equation}

The equations of motion for $\rho_x$ and $\rho_y$ can be deduced
from~\eqref{vector} at linear order $\mathcal{O}(\epsilon)$. We can further
make a variable separation as  $\rho_x(r,x)=\varphi_x(r)X(x)$ and
$\rho_y(r,x)=\varphi_y(r)Y(x)$. To satisfy the equations of motion,
$\theta$ can only be $\theta_+=\frac{\pi}{2}+2n\pi$ or
$\theta_-=-\frac{\pi}{2}+2n\pi$ with $n$ an arbitrary integer. We
then obtain the equations of motion for $\varphi_x(r)$,
$\varphi_y(r)$, $X(x)$ and $Y(x)$ as
\begin{equation}\label{eomro1}
\varphi_x \dot{X}\pm (qBx-p)\varphi_y Y=0,
\end{equation}
\begin{equation}\label{eomro2}
\varphi_x' \dot{X}\pm (qBx-p)\varphi_y' Y=0,
\end{equation}
\begin{equation}\label{eomro3}
\varphi_x''+(\frac{f'}{f}+\frac{1}{r})\varphi_x'+(\frac{q^2\mu^2}{r^2f}-\frac{m^2}{f})\varphi_x+\frac{\varphi_x}{r^2f}
[\mp (qBx-p)\frac{\dot{Y}}{X}\frac{\varphi_y}{\varphi_x}\pm qB\gamma\frac{Y}{X}\frac{\varphi_y}{\varphi_x}-(qBx-p)^2]=0,
\end{equation}
\begin{equation}\label{eomro4}
\varphi_y''+(\frac{f'}{f}+\frac{1}{r})\varphi_y'+(\frac{q^2\mu^2}{r^2f}-\frac{m^2}{f})\varphi_y+\frac{\varphi_y}{r^2f}[\frac{\ddot{Y}}{Y}\pm (qBx-p)\frac{\dot{X}}{Y}\frac{\varphi_x}{\varphi_y}
\pm (1+\gamma)qB\frac{X}{Y}\frac{\varphi_x}{\varphi_y}]=0,
\end{equation}
where the prime and dot denote the derivatives with respect to $r$
and $x$, respectively. Here and below the upper signs correspond to
the $\theta_+$ case and the lower to the $\theta_-$ case. To
satisfy~\eqref{eomro1} and ~\eqref{eomro2}, one should impose the
constraint
\begin{equation}\label{condition1}
\varphi_y=c\varphi_x, \   \ \dot{X}\pm c(qBx-p)Y=0,
\end{equation}
with $c$ a real constant.  We can see that only two of the four functions are independent. Substituting~\eqref{condition1} into the remaining equations, we can find the
following three equations
\begin{equation}\label{eomroa}
\varphi_x''+(\frac{f'}{f}+\frac{1}{r})\varphi_x'+(\frac{q^2\mu^2-\mathcal{E}}{r^2f}-\frac{m^2}{f})\varphi_x=0,
\end{equation}
\begin{equation}\label{eomrob}
-\ddot{X}\mp c(1+\gamma)qBY+(qBx-p)^2X=\mathcal{E}X,
\end{equation}
\begin{equation}\label{eomroc}
-\ddot{Y}\mp \frac{(1+\gamma)qB}{c}X+(qBx-p)^2Y=\mathcal{E}Y,
\end{equation}
where $\mathcal{E}$ is a constant. It is clear that~\eqref{eomrob}
and~\eqref{eomroc} are the same as corresponding ones in the black hole case discussed
in ref.~\cite{Cai:2013pda}. For the sake of brevity, we therefore just list the
main results in this paper. More details can be found in
ref.~\cite{Cai:2013pda}. By defining a new function as
\begin{equation}\label{psi}
\psi(x)=X(x)-Y(x),
\end{equation}
one gets its equation from~\eqref{eomrob} and~\eqref{eomroc} as
\begin{equation}\label{eompsi}
\ddot{\psi}(x)+[\mathcal{E}\mp c(1+\gamma)qB-(qBx-p)^2]\psi(x)=0,
\end{equation}
where we have imposed the condition $c^2=1$~\cite{Cai:2013pda}. This eigenvalue equation can be solved exactly with the eigenvalue function
\begin{equation}\label{solutionpsi}
\psi_n(x)=N_n e^{-\frac{1}{2}|qB|(x-\frac{p}{qB})^2}H_n(\sqrt{|qB|}(x-\frac{p}{qB})),
\end{equation}
and the corresponding eigenvalue
\begin{equation}\label{eigenvalues}
\mathcal{E}_n=(2n+1)|qB|\pm c(1+\gamma)qB,
\end{equation}
where $H_n$ is Hermite function, $N_n$ is a normalization constant
and $n$ is a non-negative integer labeling  each ``energy level''.

The solutions of $\varphi_x$ and $\varphi_y$ corresponding to the
eigenvalue $\mathcal{E}_n$, denoted by $\varphi_{xn}$ and
$\varphi_{yn}$, can be obtained by solving the equation of
motion~\eqref{eomroa} with $\mathcal{E}_n$ given
in~\eqref{eigenvalues}. So far, we have recovered the Landau levels.
As we arrive at the transition point, we should encounter a
marginally stable mode at the critical point. Those solutions
obtained from~\eqref{eomroa} just correspond to the marginally
stable states.

One can see from~\eqref{eomroa} that the effective mass of $\rho_x$ is given by
\begin{equation}\label{mass}
m_{\rm eff}^2=m^2+\frac{\mathcal{E}_n-q^2\mu^2}{r^2}=m^2+\frac{(2n+1)|qB|\pm c(1+\gamma)qB-q^2\mu^2}{r^2},
\end{equation}
which is clearly corrected by the magnetic field $B$. The appearance
of magnetic field can increase or decrease the effective mass, thus
will hinder or enhance the transition from the normal phase to the
condensed phase. In what follows, we consider the solution with the
lowest Landau level with $n=0$, which reads
\begin{equation}\label{solution1}
\begin{split}
\mathcal{E}_0^L=-|\gamma qB|,\\
X_0^L(x;p)=\frac{N_0}{2}e^{-\frac{|qB|}{2}(x-\frac{p}{qB})^2}=-Y_0^L(x;p).
\end{split}
\end{equation}
The general falloff of $\varphi_x$ near the boundary $r\rightarrow\infty$  behaves as
\begin{equation} \label{boundary}
\varphi_x=\frac{{{\rho_x}_-}}{{r^{{\Delta}_-}}}+\frac{{{\rho_x}_+}}{{r^{{\Delta}_+}}}+\ldots,
\end{equation}
with ${\Delta}_\pm=1\pm\sqrt{1+m^2}$.~\footnote{The $m^2$ has a
lower bound as $m^2=-1$ with ${\Delta}_+={\Delta}_-=1$. In such
a case, there is a logarithmic term in the asymptotical expansion. We
treat such a term as the source which is set to be vanishing to
avoid the instability induced by this term.} According to the
AdS/CFT dictionary, up to a normalization, ${\rho_x}_{-}$ and
${\rho_x}_{+}$ timing the spatial dependent part are regarded as the
source and the expectation value of the $x$  component of the dual
vector operator denoted as $\hat{J^x}$ in the boundary theory,
respectively. Similar statement holds for $\varphi_y$ corresponding
to the $y$ component of the dual vector operator $\hat{J^y}$. Since
we want the U(1) symmetry to be broken spontaneously, we turn off
the source term, namely take ${\rho_x}_{-}=0$.
\subsection{Phase diagram}
\label{sect:diagram}
We are interested in how the applied magnetic field can influence
the critical point from the normal phase to the condensed phase. As
we can see from~\eqref{mass}, the effective mass of the charged
vector field $\rho_\mu$  in the lowest Landau level $n=0$ depends on
the magnetic field $B$ and the non-minimal coupling parameter
$\gamma$ as
\begin{equation}\label{lowestmass}
m_{\rm eff}^{2}=m^2-\frac{|\gamma qB|+q^2\mu^2}{r^2}.
\end{equation}
It is clear that the increase of the magnetic field $B$ decreases
the effective mass and thus tends to induce the phase transition.
The magnetic field plays the same role as the chemical potential
$\mu$.

We introduce a new coordinate $z=r_0/r$, then the equation of motion~\eqref{eomroa} can be rewritten as~\footnote{For brevity, we also use prime to denote the derivative with respect to $z$. But the meaning of the derivative in the text is clear and will not be confused.}
\begin{equation}\label{eomroa2}
\varphi_x''(z)-\frac{1+3z^4}{z(1-z^4)}\varphi_x'(z)-[\frac{m^2}{z^2(1-z^4)}-\frac{\Lambda}{1-z^4}]\varphi_x(z)=0,
\end{equation}
where $\Lambda=\frac{q^2\mu^2+|\gamma qB|}{r_0^2}$.
\begin{figure}[h]
\centering
\includegraphics[scale=1.1]{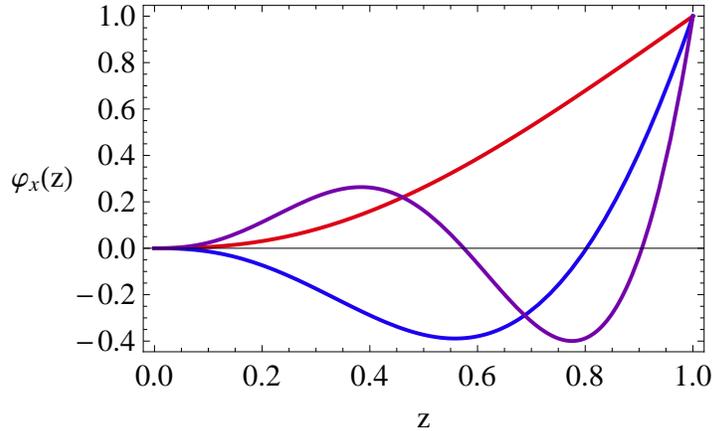}\caption{\label{modes} Three marginally stable modes of the charged vector field. The three curves from top to down correspond to $\Lambda_0\simeq7.7562$ (red), $\Lambda_1\simeq27.992$ (blue) and $\Lambda_2\simeq59.601$ (purple), respectively. We here take $m^2=5/4$. We have normalized the value of $\varphi_x$ at the tip to be one.}
\end{figure}
We further define a new function for numerical convenience
\begin{equation}\label{FF}
\varphi_x(z)=(\frac{z}{r_0})^{\triangle_-}F(z),
\end{equation}
then the equation of motion for $F(z)$ is given by
\begin{equation}\label{eomF}
\begin{split}
F''(z)-\frac{1}{z}(\frac{1+3z^4}{1-z^4}-2\triangle_-)F'(z)-[\frac{m^2+4\triangle_-}{z^2(1-z^4)}-\frac{{\triangle_-}({\triangle_-}+2)}{z^2}-\frac{\Lambda}{1-z^4}]F(z)=0.
\end{split}
\end{equation}

To solve such a second order equation, we should impose suitable
boundary conditions. Remember that the leading term of $F$ near the
boundary $z\rightarrow0$ gives the source term which is set to be zero, so
one has the condition $F(z=0)=0$ at the boundary. We impose the
regularity condition at the tip $z=1$. Thanks to the linearity
of~\eqref{eomF}, we take $F(z=1)=1$ in our numerical
calculation without loss of generality. It is clear
that~\eqref{eomF} depends on two parameters $m^2$ and $\Lambda$. For
a given $m^2$, only for certain values of $\Lambda$ can the boundary
conditions be fulfilled.

\begin{figure}[h]
\centering
\includegraphics[scale=1.1]{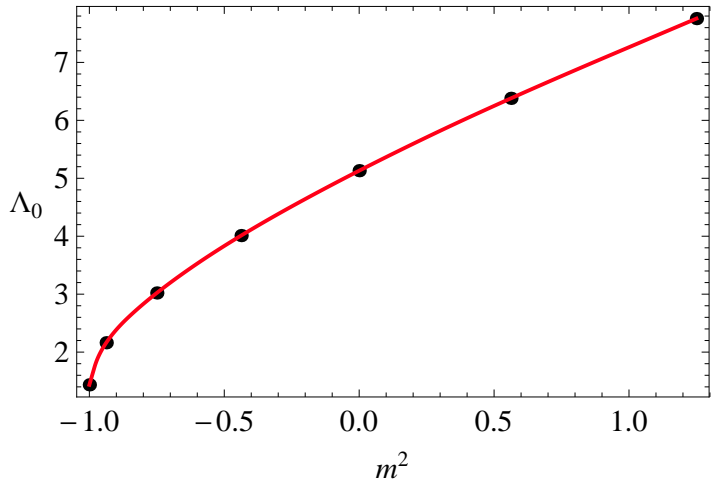}\caption{\label{eigenvalue}
The critical value of $\Lambda_0$  with respect to $m^2$ of the
vector field. The black points are obtained by solving the
equation~\eqref{eomF} numerically by use of shooting method.}
\end{figure}

Solving~\eqref{eomF} by shooting method, we present the three
lowest-lying marginally stable modes in figure~\ref{modes} for $m^2=5/4$.
 The three modes are in the sequence $\Lambda_0 < \Lambda_1 < \Lambda_2$.
 The mode with $\Lambda_0\simeq7.7562$ (the red curve) is regarded as a mode with node $n=0$
 since the curve has no intersecting points with horizontal axis except at the origin $z=0$.
  Following such terminology, the blue curve corresponding to $\Lambda_1\simeq27.992$ is
  considered as a mode with node $n=1$ and the purple one corresponding to $\Lambda_2\simeq59.601$
  as a mode with node $n=2$. Due to the radial oscillations in $z$-direction of $\varphi_x(z)$,
  the latter two modes are therefore thought to be less stable than the first one.
  Thus, the lowest value $\Lambda_0$ just gives the critical value above which the AdS soliton
  is unstable to developing a vector ``hair". We plot the value of $\Lambda_0$ for different squared mass
  of the vector field in figure~\ref{eigenvalue}, from which one can see that $\Lambda_0$ increases as we increase the squared mass.

It should be stressed here that it is the combination
$\Lambda_0=\frac{q^2\mu^2+|\gamma qB|}{r_0^2}$ that determines the
phase transition point of the system. More specifically, turning
off the chemical potential $\mu$, we can see that only the magnetic
field itself can also induce the instability. This result is
totally  different from the case in the holographic s-wave
superconductor/insulator model~\cite{Cai:2011tm}, where the presence
of the magnetic field makes the normal phase more stable rather than
unstable. The result here is reminiscent of the QCD vacuum
instability induced by strong magnetic field to spontaneously
developing the $\rho$-meson condensate. Just as pointed out in the
finite temperature case~\cite{Cai:2013pda}, this is due to the
non-minimal coupling of the vector field $\rho^\mu$ to the gauge
field $A_\mu$ in the last term in~\eqref{action}. The reader would
be familiar with such a term used to describe the coupling of
magnetic moment for charged vector particles to a background
magnetic field in the literature, see
refs.~\cite{Young:1963zza,Djukanovic:2005ag} for example.

%

\section{Einstein-Yang-Mills model}
\label{sect:nonabeliabmodel}
The black hole instability induced by an applied non-Abelian
magnetic field in the SU(2) p-wave model was studied in
refs.~\cite{Wong:2013rda,Ammon:2011je}. In this section, we will
study the AdS soliton instability induced by the non-Abelian magnetic
field.

The five-dimensional Einstein-SU(2) Yang-Mills theory with a
negative cosmological constant can be written
as~\cite{Gubser:2008wv}
\begin{equation}\label{actionsu2}
S =\int d^5 x
\sqrt{-g}[\frac{1}{2\kappa^2}(\mathcal{R}+\frac{12}{L^2})-\frac{1}{4\hat{g}^2}
F^a_{\mu\nu} F^{a\mu \nu}],
\end{equation}
where $\hat{g}$ is the Yang-Mills coupling constant. The SU(2)
Yang-Mills field strength is given
by~\footnote{$\mu,\nu=(t,r,x,y,\eta)$ denote the indices of
spacetime and $a,b,c=(1,2,3)$ are the indices of the SU(2) group
generators $\tau^a=\sigma^a/2i$, where $\sigma^a$ are Pauli
matrices.}
\begin{equation}
F^a_{\mu\nu}=\partial_\mu A^a_\nu-\partial_\nu A^a_\mu + \epsilon^{abc}A^b_\mu A^c_\nu,
\end{equation}
where $\epsilon^{abc}$ is the totally antisymmetric tensor with $\epsilon^{123}=+1$. The gauge field is given by $A=A^a_{\mu}\tau^adx^{\mu}$. The equation of motion for $A^a_\mu$ reads
\begin{equation}\label{eomsu2}
\nabla_\mu F^{a\mu\nu}+\epsilon^{abc}A^b_\mu F^{c\mu\nu}=0.
\end{equation}
We work in the probe limit and treat marginally stable perturbations as probe into the AdS soliton~\eqref{metric}.
\subsection{Adding a constant magnetic field}
\label{sect:ansatzBsu2}

We turn on a background non-Abelian magnetic field $B$ and also allow for a non-Abelian chemical potential in the dual system. Both of them come from the $\tau^3$ component of the Lie algebra.  We take the following ansatz
\begin{equation}\label{ansatzsu2}
\begin{split}
A^1_\mu dx^\mu&=[\epsilon a^1_x(r,x,y)+\mathcal{O}(\epsilon^3)]dx+[\epsilon a^1_y(r,x,y)+\mathcal{O}(\epsilon^3)]dy,\\
A^2_\mu dx^\mu&=[\epsilon a^2_x(r,x,y)+\mathcal{O}(\epsilon^3)]dx+[\epsilon a^2_y(r,x,y)+\mathcal{O}(\epsilon^3)]dy,\\
A^3_\mu dx^\mu&=[\phi(r)+\mathcal{O}(\epsilon^2)]dt+[Bx+\mathcal{O}(\epsilon^2)]dy,\\
\end{split}
\end{equation}
with $\epsilon$ the small deviation parameter.
Substituting above ansatz into~\eqref{eomsu2}, we obtain the equation of motion for $\phi(r)$ at leading order, which is just the same as~\eqref{eomphi} with the solution given in~\eqref{phi}.
At linear order $\mathcal{O}(\epsilon)$, the equations of motion turn out to be
\begin{equation}\label{eomscu2}
\partial_xV_x+(\partial_y-iBx)V_y=0,
\end{equation}
\begin{equation}\label{eomsdu2}
\partial_x\partial_rV_x+(\partial_y-iBx)\partial_rV_y=0,
\end{equation}
\begin{equation}\label{eomsau2}
\partial_r^2V_x+(\frac{f'}{f}+\frac{1}{r})\partial_rV_x+\frac{1}{r^2f}[(\partial_y^2-i2Bx\partial_y+\mu^2-B^2x^2)V_x+(-\partial_x\partial_y+iBx\partial_x-iB)V_y]=0,
\end{equation}
\begin{equation}\label{eomsbu2}
\partial_r^2V_y+(\frac{f'}{f}+\frac{1}{r})\partial_rV_y+\frac{1}{r^2f}[(-\partial_x\partial_y+iBx\partial_x+i2B)V_x+(\partial_x^2+\mu^2)V_y]=0,
\end{equation}
where $V_x=a_x^1-ia_x^2$ and $V_y=a_y^1-ia_y^2$.

In order to solve above four equations of motion, we further make a variable separation
\begin{equation}\label{eomscu2}
V_x(r,x,y)=\tilde{\varphi}_x(r)\tilde{X}(x)e^{ipy},\quad V_y(r,x,y)=\tilde{\varphi}_y(r)\tilde{Y}(x)e^{ipy}e^{i\tilde{\theta}},
\end{equation}
where $\tilde{\varphi}_x$, $\tilde{\varphi}_y$, $\tilde{X}$ and $\tilde{Y}$ are all real functions and $p$ is a real constant. We also introduce a constant $\tilde{\theta}$ describing the phase difference between $V_x$ and $V_y$. Further analysis shows that $\tilde{\theta}$ can only be $\tilde{\theta}_+=\frac{\pi}{2}+2n\pi$ or $\tilde{\theta}_-=-\frac{\pi}{2}+2n\pi$ with $n$ an arbitrary integer. Then we obtain the following equations of motion
\begin{equation}\label{eomro1su2}
\tilde{\varphi}_x \dot{\tilde{X}}\pm (Bx-p)\tilde{\varphi}_y \tilde{Y}=0,
\end{equation}
\begin{equation}\label{eomro2su2}
\tilde{\varphi}_x' \dot{\tilde{X}}\pm (Bx-p)\tilde{\varphi}_y' \tilde{Y}=0,
\end{equation}
\begin{equation}\label{eomro3su2}
\tilde{\varphi}_x''+(\frac{f'}{f}+\frac{1}{r})\tilde{\varphi}_x'+\frac{\mu^2}{r^2f}\tilde{\varphi}_x+\frac{\tilde{\varphi}_x}{r^2f}
[\mp (Bx-p)\frac{\dot{\tilde{Y}}}{\tilde{X}}\frac{\tilde{\varphi}_y}{\tilde{\varphi}_x}\pm B\frac{\tilde{Y}}{\tilde{X}}\frac{\tilde{\varphi}_y}{\tilde{\varphi}_x}-(Bx-p)^2]=0,
\end{equation}
\begin{equation}\label{eomro4su2}
\tilde{\varphi}_y''+(\frac{f'}{f}+\frac{1}{r})\tilde{\varphi}_y'+\frac{\mu^2}{r^2f}\tilde{\varphi}_y+\frac{\tilde{\varphi}_y}{r^2f}[\frac{\ddot{\tilde{Y}}}{\tilde{Y}}\pm (Bx-p)
\frac{\dot{\tilde{X}}}{\tilde{Y}}\frac{\tilde{\varphi}_x}{\tilde{\varphi}_y}\pm 2B\frac{\tilde{X}}{\tilde{Y}}\frac{\tilde{\varphi}_x}{\tilde{\varphi}_y}]=0,
\end{equation}
where the prime and dot denote the derivatives with respect to $r$ and $x$, respectively. The same as in the previous section, the upper signs correspond to the $\tilde{\theta}_+$ case and the lower to the $\tilde{\theta}_-$ case. Comparing the above four equations of motion with equations~\eqref{eomro1}-\eqref{eomro4}, we can find that the two sets of equations of motion are the same if we choose parameters $q=1, m^2=0$ and $\gamma=1$ in~\eqref{eomro1}-\eqref{eomro4}. Thus, the results obtained in the previous section can be used here by simply restricting the model parameters to $q=1, m^2=0$ and $\gamma=1$. In this sense, the complex vector field model can be thought of
as a generalization of the SU(2) p-wave model.
\subsection{Phase diagram}
\label{sect:diagramsu2}
In the presence of constant magnetic field and chemical potential, the field $\varphi_x$~\footnote{Since the non-Abelian p-wave model in our ansatz is the same as the Abelian p-wave model with parameters $q=1, m^2=0$ and $\gamma=1$, in what follows, we shall omit the tilde in all functions in this non-Abelian model for clarity.} acquires an effective mass in the lowest Landau level $n=0$ as
\begin{equation}\label{lowmass}
m_{\rm eff}^2=\frac{\mathcal{E}_0-\mu^2}{r^2}=-\frac{|B|+\mu^2}{r^2}.
\end{equation}
It is clear that the increase of the magnetic field $B$ decreases the effective mass and thus tends to induce the instability. For given chemical potential $\mu$, as we increase the magnetic field $B$ to the critical value $\Lambda_0 r_0^2-\mu^2$ with $\Lambda_0=(\mu^2+|B|)/r_0^2\simeq 5.1313$, the AdS soliton background will become unstable to developing non-trivial vector ``hair". It is clear here that even without the chemical potential, the magnetic field will induce the vector condensate and the AdS soliton instability. In addition, note that the critical value $\Lambda_0$ is comparable to the critical magnetic field found in the black hole background~\cite{Ammon:2011je}.
\begin{figure}[h]
\centering
\includegraphics[scale=1.1]{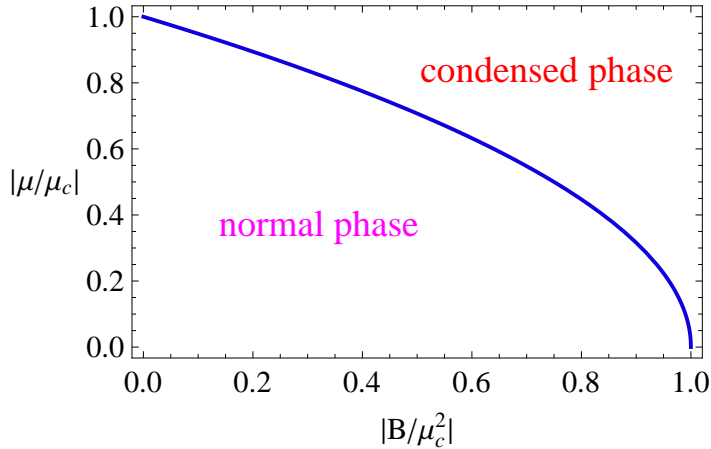}\caption{\label{phasemuB}  The value of chemical potential from the normal phase to the condensed phase as a function of magnetic field. It corresponds to the lowest Landau level with $\Lambda_0\simeq 5.1313$. $\mu_c=\sqrt{\Lambda_0}r_0$ is the critical chemical potential in the case without the applied magnetic field.}
\end{figure}

The $(\mu, B)$ phase diagram for the lowest Landau level with $\Lambda_0\simeq 5.1313$ is drawn in figure~\ref{phasemuB}. In order to determine which side of the phase boundary is the condensed phase, we take a look at the equation~\eqref{lowmass} which suggests that the magnetic field decreases the effective mass. As one increases the strength of magnetic field at a fixed chemical potential, the normal phase will become unstable for sufficiently large magnetic field. Therefore, the region in the upper right in figure~\ref{phasemuB} corresponds to the condensed phase. In addition, let us mention here that the $(\mu, B)$ phase diagram for a fixed $m^2$ for the Abelian p-wave model discussed in the previous section should be extremely similar to figure~\ref{phasemuB}.

\section{Vortex lattice solution}
\label{sect:vortex}
Following  ref.~\cite{Cai:2013pda}, let us now construct the vortex lattice solution for the Abelian and the non-Abelian p-wave models. We consider the case with the lowest Landau level $n=0$.
Since the eigenvalue $\mathcal{E}_0$ is independent of $p$, it is easy to find that a linear superposition of the solutions $e^{ipy}\varphi_{x0}(r)X_0^L(x;p)$ and $e^{ipy}\varphi_{y0}(r)Y_0^L(x;p)$ with different $p$ is also a solution of two models at linear order $\mathcal{O}(\epsilon)$. Therefore, the following two functions
\begin{equation}\label{newfunctions}
\begin{split}
W_x(r,x,y)=\varphi_{x0}(r)\sum_{\ell=-\infty}^{+\infty}c_\ell e^{ip_\ell y}X_0^L(x;p_\ell),\\
W_y(r,x,y)=ce^{i\theta_\pm}\varphi_{x0}(r)\sum_{\ell=-\infty}^{+\infty}c_\ell e^{ip_\ell y}Y_0^L(x;p_\ell),
\end{split}
\end{equation}
are solutions of the full equations of motion, where $c_\ell=e^{-i\frac{\pi a_2}{a_1^2}\ell^2}$ and $p_\ell=\frac{2\pi\sqrt{|qB|} \ell}{a_1}$ with $a_1$ and $a_2$ arbitrary constants. According to ref.~\cite{Cai:2013pda}, we can construct the vortex lattice solution
\begin{equation}\label{superposition}
\triangle W(r,x,y)\equiv W_x(r,x,y)-ce^{-i\theta_\pm}W_y(r,x,y)=\varphi_{x0}(r)\sum_{\ell=-\infty}^{+\infty}c_\ell e^{ip_\ell y}\psi_0(x;p_\ell),
\end{equation}
where $\psi_0(x;p_\ell)=N_0e^{-\frac{1}{2}|qB|(x-\frac{p_\ell}{qB})^2}$.

The solution $\triangle W(r,x,y)$ exhibits a pseudo-periodicity
\begin{equation}\label{property1}
\begin{split}
\triangle W(r,x,y)=\triangle W(r,x,y+\frac{a_1}{\sqrt{|qB|}}),\\
\triangle W(r,x+\frac{2\pi}{a_1\sqrt{|qB|}},y+\frac{a_2}{a_1\sqrt{|qB|}})=e^{\frac{i2\pi}{a_1}(\sqrt{|qB|}y+\frac{a_2}{2a_1})}\triangle W(r,x,y),
\end{split}
\end{equation}
and has a zero at
\begin{equation}\label{property2}
\textbf{x}_{\tilde{m},\tilde{n}}=(\tilde{m}+\frac{1}{2})\textbf{v}_1+(\tilde{n}+\frac{1}{2})\textbf{v}_2,
\end{equation}
where two vectors $\textbf{v}_1=\frac{a_1}{\sqrt{|qB|}}\partial_y$ and $\textbf{v}_2=\frac{2\pi}{a_1\sqrt{|qB|}}\partial_x+\frac{a_2}{a_1\sqrt{|qB|}}\partial_y$~\cite{Maeda:2009vf}. $\tilde{m}$ and $\tilde{n}$ are two integers. Remember that the corresponding results for the SU(2) model can be directly obtained by choosing the parameters $q=1, \gamma=1$ and $m^2=0$.

According to the AdS/CFT dictionary, the coefficient of the sub-leading term near the boundary $r\rightarrow\infty$ of $\triangle W(r,x,y)$ just gives the vacuum expectation value of the dual vector operator. For the Abelian p-wave model, the expectation value of the operator $\hat{J^x}$ dual to $\rho_x$ is given by the coefficient of $1/r^{{\Delta_+}}$ at boundary $r\rightarrow\infty$, the combination $J_{\pm}=\langle\hat{J^x}\pm i\hat{J^y}\rangle$ dose exhibit the vortex lattice structure which has the cores located at $\textbf{x}_{\tilde{m},\tilde{n}}$. This result is the same as that we reported in the black hole background~\cite{Cai:2013pda}. In particular, to obtain the triangular lattice we can choose the following parameters
\begin{equation}\label{triangular}
a_1=\frac{2\sqrt{\pi}}{\sqrt[4]{3}},\    \ a_2=\frac{2\pi}{\sqrt{3}}.
\end{equation}
We notice that it is the linear combinations $J_{\pm}$ which exhibit the vortex lattice solution. In particular, for $q>0$, it is $J_-$ that corresponds to the lowest Landau level, while for $q<0$, it is $J_+$. The SU(2) p-wave model corresponds to the case with $q=1$, so its vortex lattice solution is related to $J_{-}$. This further provides the evidence for the correctness of choosing $c^2=1$ in section~\ref{sect:ansatzB}.

\begin{figure}[h]
\centering
\includegraphics[scale=0.8]{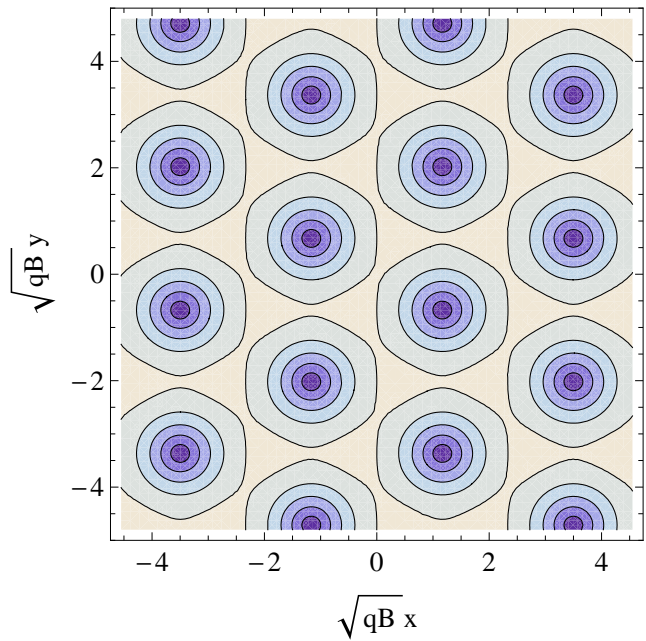}\caption{\label{lattice} The contour plot of the vortex lattice structure for the triangular lattice in $x-y$ plane. Darker shading indicates smaller values of the norm of the condensate. In particular, the condensate vanishes at the core of each vortex.}
\end{figure}

As a concrete example, the vortex lattice is presented in figure~\ref{lattice} in the $x-y$ plane perpendicular to the magnetic field for the triangular structure. For other values of $a_1$ and $a_2$, one can obtain different vortex lattice configurations. To determine the true ground state, one must go beyond the linear analysis to find the values of $a_1$ and $a_2$ which minimize the free energy of all possible lattice structures. What's more, the overall normalization of $\triangle W(r,x,y)$ can only be fixed at higher order. The calculation of higher order contributions can be done straightforwardly as in refs.~\cite{Bu:2012mq,Maeda:2009vf}. Nevertheless, it is much more involved and is not very relevant to our purpose of this paper. Thus, we shall leave it for our further study.
%

\section{Conclusion and discussion}
\label{sect:conclusion}
As a continuation of our recent studies on a holographic p-wave superconductor model in the Einstein-Maxwell-complex vector field model proposed in ref.~\cite{Cai:2013pda}, in this paper we generalized to the case with vector condensate of dual conformal field theory in confined phase, i.e., in an AdS soliton background (see the metric~\eqref{metric}).  This study is relevant to the p-wave superconductor(superfluid)/insulator phase transition as well as charged $\rho$-meson condensation in confined QCD vacuum in strong magnetic field. We focused on the behavior of the model in the presence of a constant magnetic field $B$. At linear order $\mathcal{O}(\epsilon)$, the spatial part of vector field $\rho_\mu$ can be solved analytically, which gives the Landau level of the vector field in a magnetic field background.  It was found that the effective mass of the vector field is reduced by turning on magnetic field in the lowest Landau level case. More precisely, it is the combination $\Lambda=(q^2\mu^2+|\gamma qB|)/r_0^2$ that determines the instability of our system, here $\mu$ is the chemical potential of dual field theory, while $r_0$ is related to the mass gap of the field theory. When we increase the value of $\Lambda$ to the critical one, the AdS soliton background becomes unstable to developing non-trivial vector ``hair". Interestingly, it can be seen from $\Lambda$ that the magnetic field itself can induce the instability even without chemical potential $\mu$. It is clear that the essential point for this is due to the non-minimal coupling between the vector field and the background magnetic field.

To have  further understanding, we also study the Einstein-SU(2) Yang-Mills theory in the AdS soliton background with a non-Abelian magnetic field. Interestingly, at least under our ansatz in the present paper, we found that the reduced equations of motion for the non-Abelian model are the same as those of the complex vector field model in the case with parameters $m^2=0$ and $\gamma=1$.~\footnote{The condition $q=1$ is not important, since the charge $q$ of the vector field $\rho_\mu$ can be scaled to be one in the probe limit.} Therefore, the presence of a non-Abelian magnetic field can also make the AdS soliton background unstable. In some sense, the Einstein-Maxwell-complex vector field model is a generalization of the SU(2) model to a general mass square $m^2$ and gyromagnetic ratio $\gamma$ of the vector field.

The phase diagram in terms of $B$ and $\mu$ is drawn in figure~\ref{phasemuB}. The critical chemical potential above which a condensed phase appears decreases with the applied magnetic field, which means the presence of magnetic field can enhance rather than suppress the phase transition. This result is reminiscent of the condensation of $W$-meson~\cite{Ambjorn:1988fx} and $\rho$-meson~\cite{Chernodub:2010qx,Chernodub:2011mc} in a strong magnetic field. Of course, this result is opposite to the well known behavior reported in ordinary superconductors as well as holographic s-wave superconductor/insulator model~\cite{Cai:2011tm}, there the presence of an applied magnetic field will suppress the superconducting phase transition. Remember that the action~\eqref{action} describes a p-wave model, and that for the p-wave superconductors, theoretical studies~\cite{PhysRevLett.58.1482,Rasolt:1992zz,Klimenko:2012qi,Olesen:2013nca} as well as experimental evidences~\cite{levy2005,uji2010} suggest the possibility that the magnetic field can induce superconductivity in some cases, in particular, for the ferromagnetic superconductors. Therefore our results are understandable.

In the present study, we limited ourselves to the probe approximation by neglecting the back reaction of matter fields on the bulk AdS soliton geometry, which may work only near the critical point with continuous phase transition. Although it does be able to reveal some significant properties, something would be lost in this approximation, especially for the whole phase structure, see ref.~\cite{Cai:2013wma} as an example. Indeed, in the black hole background case, going beyond the probe approximation, we found a rich phase structure in this Einstein-Maxwell-complex vector field model~\cite{Cai:2013aca}. Depending on mass square $m^2$ and charge $q$, the phase transition can be zeroth order, first order or second order. In addition, there also exists the so-called ``retrograde condensation". The behavior of entanglement entropy in this model has also been studied in ref.~\cite{Li:2013rhw}. In some sense, although the Einstein-Maxwell-complex vector model looks to be a generalization of the SU(2) p-wave model, it can have much rich phase structure and distinguished properties compared to the latter. Therefore, it is interesting to go beyond the probe limit in the soliton background. Our ongoing work shows that this model has  much more interesting behaviors in the soliton background.

The Einstein-Maxwell-complex vector field model described by the action (\ref{action}) is just a phenomenological model. At the moment it is not sure whether this model can be embedded into some effective theory of string/M theory. If the answer is yes, it is certainly of great interest to understand those interesting properties of the model from the dual field theory side. This must be quite helpful to understand
some high temperature superconductors and QCD vacuum in strong magnetic field.

\section*{Acknowledgements}

This work was supported in part by the National Natural Science Foundation of China (No.10821504, No.11035008, No.11205226 and No.11305235), and in part by the Ministry of Science and Technology of China under Grant No.2010CB833004.

\appendix


\begin{thebibliography}{99}


\bibitem{Maldacena:1997re}
  J.~M.~Maldacena,
  ``The large N limit of superconformal field theories and supergravity,''
  Adv.\ Theor.\ Math.\ Phys.\  {\bf 2}, 231 (1998)
  [Int.\ J.\ Theor.\ Phys.\  {\bf 38}, 1113 (1999)]
  [arXiv:hep-th/9711200].
\bibitem{Gubser:1998bc}
  S.~S.~Gubser, I.~R.~Klebanov and A.~M.~Polyakov,
  ``Gauge theory correlators from non-critical string theory,''
  Phys.\ Lett.\  B {\bf 428}, 105 (1998)
  [arXiv:hep-th/9802109].
\bibitem{Witten:1998qj}
  E.~Witten,
  ``Anti-de Sitter space and holography,''
  Adv.\ Theor.\ Math.\ Phys.\  {\bf 2}, 253 (1998)
  [arXiv:hep-th/9802150].

\bibitem{Hartnoll:2009sz}
  S.~A.~Hartnoll,
  ``Lectures on holographic methods for condensed matter physics,''
  Class.\ Quant.\ Grav.\  {\bf 26}, 224002 (2009)
  [arXiv:0903.3246 [hep-th]].

\bibitem{Herzog:2009xv}
  C.~P.~Herzog,
  ``Lectures on Holographic Superfluidity and Superconductivity,''
  J.\ Phys.\ A {\bf 42}, 343001 (2009)
  [arXiv:0904.1975 [hep-th]].

\bibitem{McGreevy:2009xe}
  J.~McGreevy,
  ``Holographic duality with a view toward many-body physics,''
  Adv.\ High Energy Phys.\  {\bf 2010}, 723105 (2010)
  [arXiv:0909.0518 [hep-th]].

\bibitem{Hartnoll:2008vx}
S.~A.~Hartnoll, C.~P.~Herzog and G.~T.~Horowitz,
``Building a Holographic Superconductor,''
Phys.\ Rev.\ Lett.\  {\bf 101}, 031601 (2008)
[arXiv:0803.3295 [hep-th]].

\bibitem{Hartnoll:2008kx}
  S.~A.~Hartnoll, C.~P.~Herzog and G.~T.~Horowitz,
  ``Holographic Superconductors,''  JHEP {\bf 0812}, 015 (2008)  [arXiv:0810.1563 [hep-th]].


\bibitem{Gubser:2008wv}
  S.~S.~Gubser and S.~S.~Pufu,
  ``The Gravity dual of a p-wave superconductor,''  JHEP {\bf 0811}, 033 (2008)  [arXiv:0805.2960 [hep-th]].

\bibitem{Aprile:2010ge}
  F.~Aprile, D.~Rodriguez-Gomez and J.~G.~Russo,
  ``p-wave Holographic Superconductors and five-dimensional gauged Supergravity,''
  JHEP {\bf 1101}, 056 (2011)
  [arXiv:1011.2172 [hep-th]].

\bibitem{Cai:2013pda}
  R.~-G.~Cai, S.~He, L.~Li and L.~-F.~Li,
  ``A Holographic Study on Vector Condensate Induced by a Magnetic Field,''
  arXiv:1309.2098 [hep-th].

\bibitem{Cai:2013aca}
  R.~-G.~Cai, L.~Li and L.~-F.~Li,
  ``A Holographic P-wave Superconductor Model,''
  arXiv:1309.4877 [hep-th].

\bibitem{Chen:2010mk}
  J.~-W.~Chen, Y.~-J.~Kao, D.~Maity, W.~-Y.~Wen and C.~-P.~Yeh,
  ``Towards A Holographic Model of D-Wave Superconductors,''
  Phys.\ Rev.\ D {\bf 81}, 106008 (2010)
  [arXiv:1003.2991 [hep-th]].

\bibitem{Benini:2010pr}
  F.~Benini, C.~P.~Herzog, R.~Rahman and A.~Yarom,
  ``Gauge gravity duality for d-wave superconductors: prospects and challenges,''
  JHEP {\bf 1011}, 137 (2010)
  [arXiv:1007.1981 [hep-th]].




\bibitem{Kharzeev:2012ph}
D.~E.~Kharzeev, K.~Landsteiner, A.~Schmitt and H.~-U.~Yee,
```Strongly interacting matter in magnetic fields': an overview,''
Lect.\ Notes Phys.\  {\bf 871}, 1 (2013)
[arXiv:1211.6245 [hep-ph]].

\bibitem{Chernodub:2010qx}
  M.~N.~Chernodub,
  ``Superconductivity of QCD vacuum in strong magnetic field,''
  Phys.\ Rev.\ D {\bf 82}, 085011 (2010)
  [arXiv:1008.1055 [hep-ph]].

\bibitem{Chernodub:2011mc}
  M.~N.~Chernodub,
  ``Spontaneous electromagnetic superconductivity of vacuum in strong magnetic field: evidence from the Nambu--Jona-Lasinio model,''
  Phys.\ Rev.\ Lett.\  {\bf 106}, 142003 (2011)
  [arXiv:1101.0117 [hep-ph]].

\bibitem{Chernodub:2011gs}
  M.~N.~Chernodub, J.~Van Doorsselaere and H.~Verschelde,
  ``Electromagnetically superconducting phase of vacuum in strong magnetic field: structure of superconductor and superfluid vortex lattices in the ground state,''
  Phys.\ Rev.\ D {\bf 85}, 045002 (2012)
  [arXiv:1111.4401 [hep-ph]].



\bibitem{Callebaut:2011ab}
  N.~Callebaut, D.~Dudal and H.~Verschelde,
  ``Holographic rho mesons in an external magnetic field,''
  JHEP {\bf 1303}, 033 (2013)
  [arXiv:1105.2217 [hep-th]].

\bibitem{Callebaut:2013ria}
  N.~Callebaut and D.~Dudal,
  ``On the transition temperature(s) of magnetized two-flavour holographic QCD,''
  arXiv:1303.5674 [hep-th].

\bibitem{Callebaut:2013wba}
  N.~Callebaut and D.~Dudal,
  ``A magnetic instability of the non-Abelian Sakai-Sugimoto model,''
  arXiv:1309.5042 [hep-th].

\bibitem{Bu:2012mq}
  Y.~-Y.~Bu, J.~Erdmenger, J.~P.~Shock and M.~Strydom,
  ``Magnetic field induced lattice ground states from holography,''
  JHEP {\bf 1303}, 165 (2013)
  [arXiv:1210.6669 [hep-th]].

\bibitem{Wong:2013rda}
  K.~Wong,
  ``A non-abelian vortex lattice in strongly-coupled systems,''
  arXiv:1307.7839 [hep-th].


\bibitem{Nishioka:2009zj}
  T.~Nishioka, S.~Ryu and T.~Takayanagi,
  ``Holographic Superconductor/Insulator Transition at Zero Temperature,''
  JHEP {\bf 1003}, 131 (2010)
  [arXiv:0911.0962 [hep-th]].

\bibitem{Horowitz:2010jq}
  G.~T.~Horowitz and B.~Way,
  ``Complete Phase Diagrams for a Holographic Superconductor/Insulator System,''
  JHEP {\bf 1011}, 011 (2010)
  [arXiv:1007.3714 [hep-th]].

\bibitem{Peng:2011gh}
  Y.~Peng, Q.~Pan and B.~Wang,
  ``Various types of phase transitions in the AdS soliton background,''
  Phys.\ Lett.\ B {\bf 699}, 383 (2011)
  [arXiv:1104.2478 [hep-th]].


\bibitem{Cai:2012es}
  R.~-G.~Cai, S.~He, L.~Li and L.~-F.~Li,
  ``Entanglement Entropy and Wilson Loop in St\'{u}ckelberg Holographic Insulator/Superconductor Model,''
  JHEP {\bf 1210}, 107 (2012)
  [arXiv:1209.1019 [hep-th]].


\bibitem{Akhavan:2010bf}
  A.~Akhavan and M.~Alishahiha,
  ``P-Wave Holographic Insulator/Superconductor Phase Transition,''
  Phys.\ Rev.\ D {\bf 83}, 086003 (2011)
  [arXiv:1011.6158 [hep-th]].


\bibitem{Cai:2013oma}
  R.~-G.~Cai, L.~Li, L.~-F.~Li and R.~-K.~Su,
  ``Entanglement Entropy in Holographic P-Wave Superconductor/Insulator Model,''
  JHEP {\bf 1306}, 063 (2013)
  [arXiv:1303.4828 [hep-th]].

\bibitem{Albash:2008eh}
  T.~Albash and C.~V.~Johnson,
  ``A Holographic Superconductor in an External Magnetic Field,''
  JHEP {\bf 0809}, 121 (2008)
  [arXiv:0804.3466 [hep-th]].

\bibitem{Cai:2011tm}
  R.~-G.~Cai, L.~Li, H.~-Q.~Zhang and Y.~-L.~Zhang,
  ``Magnetic Field Effect on the Phase Transition in AdS Soliton Spacetime,''
  Phys.\ Rev.\ D {\bf 84}, 126008 (2011)
  [arXiv:1109.5885 [hep-th]].


\bibitem{Young:1963zza}
  J.~A.~Young and S.~A.~Bludman,
  ``Electromagnetic Properties of a Charged Vector Meson,''
  Phys.\ Rev.\  {\bf 131}, 2326 (1963).

\bibitem{Djukanovic:2005ag}
  D.~Djukanovic, M.~R.~Schindler, J.~Gegelia and S.~Scherer,
  ``Quantum electrodynamics for vector mesons,''
  Phys.\ Rev.\ Lett.\  {\bf 95}, 012001 (2005)
  [hep-ph/0505180].

\bibitem{Ammon:2011je}
  M.~Ammon, J.~Erdmenger, P.~Kerner and M.~Strydom,
``Black Hole Instability Induced by a Magnetic Field,''
  Phys.\ Lett.\ B {\bf 706}, 94 (2011)
  [arXiv:1106.4551 [hep-th]].

\bibitem{Maeda:2009vf}
  K.~Maeda, M.~Natsuume and T.~Okamura,
  ``Vortex lattice for a holographic superconductor,''
  Phys.\ Rev.\ D {\bf 81}, 026002 (2010)
  [arXiv:0910.4475 [hep-th]].


\bibitem{Ambjorn:1988fx}
  J.~Ambjorn and P.~Olesen,
  ``Antiscreening Of Large Magnetic Fields By Vector Bosons,''
  Phys.\ Lett.\ B {\bf 214}, 565 (1988).


\bibitem{PhysRevLett.58.1482}
Rasolt.~Mark,
``Superconductivity in high magnetic fields,''
Phys.\ Rev.\ Lett.\  {\bf 58}, 1482 (1987).

\bibitem{Rasolt:1992zz}
M.~Rasolt and Z.~Tesanovic,
``Theoretical aspects of superconductivity in very high magnetic fields,''
Rev.\ Mod.\ Phys.\  {\bf 64}, 709 (1992).

\bibitem{Klimenko:2012qi}
  K.~G.~Klimenko, R.~N.~Zhokhov and V.~C.~.Zhukovsky,
  ``Superconductivity phenomenon induced by external in-plane magnetic field in (2+1)-dimensional Gross-Neveu type model,''
  Mod.\ Phys.\ Lett.\ A {\bf 28}, 1350096 (2013)
  [arXiv:1211.0148 [hep-th]].

\bibitem{Olesen:2013nca}
  P.~Olesen,
  ``Anti-screening ferromagnetic superconductivity,''
  arXiv:1311.4519 [hep-th].

\bibitem{levy2005}
 F. Levy, I. Sheikin, B. Grenier, A. Huxley,
``Magnetic Field-induced Superconductivity in the Ferromagnet URhGe,''
 Science \textbf{309} 1343 (2005).

\bibitem{uji2010}
S. Uji, H. Shinagawa, T. Terashima, T. Yakabe, Y. Teral, M. Tokumoto, A. Kobayashi, H. Tanaka, H. Kobayashi,
``Magnetic-field-induced superconductivity in a two-dimensional organic conductor,''
Nature \textbf{410} 908 (2010).



\bibitem{Cai:2013wma}
  R.~-G.~Cai, L.~Li, L.~-F.~Li and Y.~-Q.~Wang,
  ``Competition and Coexistence of Order Parameters in Holographic Multi-Band Superconductors,''
  JHEP {\bf 1309}, 074 (2013)
  [arXiv:1307.2768 [hep-th]].


\bibitem{Li:2013rhw}
  L.~-F.~Li, R.~-G.~Cai, L.~Li and C.~Shen,
  ``Entanglement entropy in a holographic p-wave superconductor model,''
  arXiv:1310.6239 [hep-th].

\baselineskip 12pt










\end{thebibliography}
\end{document}